\documentclass[12pt]{iopart}
\usepackage{epsfig}
\usepackage{iopams}

\newcommand{\be}{\begin{equation}}
\newcommand{\ee}{\end{equation}}

\newcommand{\ba}{\begin{array}}
\newcommand{\ea}{\end{array}}
\newcommand{\bea}{\begin{eqnarray}}
\newcommand{\eea}{\end{eqnarray}}

\newcommand{\g}{\mathfrak{g}}
\newcommand{\glr}{sl(2, \mathbb{R})}
\newcommand{\Glr}{SL(2, \mathbb{R})}

\newcommand{\nn}{\nonumber \\}
\newcommand{\prw}{pr \vec{v}_Q }

\newcommand{\px}{D_x} 
\newcommand{\py}{D_y}

\begin{document}

\title{Surfaces immersed in Lie algebras associated with elliptic integrals}
\author{ A M Grundland$^1$ $^2$ and S Post$^1$}

\address{$^1$ Centre de Recherches Math\'ematiques. Universit\'e de Montr\'eal. Montr\'eal CP6128 (QC) H3C 3J7, Canada}
\address{$^2$ Department of Mathematics and Computer Sciences, Universit\'e du Quebec, Trois-Riviers. CP500 (QC)G9A 5H7, Canada}
\ead{grundlan@crm.umontreal.ca, post@crm.umontreal.ca}

\begin{abstract} The main aim of this paper is to study soliton surfaces immersed in Lie algebras associated with ordinary differential equations (ODE's) for elliptic functions. That is, given a linear spectral problem for such an ODE in matrix Lax representation, we search for the most general solution of the wave function which satisfies the linear spectral problem. These solutions allow for the explicit construction of soliton surfaces by the Fokas-Gel'fand formula for immersion, as formulated in \cite{GrundPost2011} which is based on the formalism of generalized vector fields and their prolongation structures. The problem has been reduced to examining three types of symmetries, namely, a conformal symmetry in the spectral parameter (known as the Sym-Tafel formula), gauge transformations of the wave function and generalized symmetries of the  associated integrable ODE. The paper contains a detailed explanation of the immersion theory of surfaces in Lie algebras in connection with ODE's as well as an exposition of the main tools used to study their geometric characteristics. Several examples of the Jacobian and $\mathcal{P}$-Weierstrass elliptic functions are included as illustrations of the theoretical results.  
\end{abstract}
\maketitle

\section{Introduction}\label{intro}
The language of group theoretical methods for investigating soliton surfaces associated with integrable models is a very useful and adequate tool for studying the main features of various problems appearing in diverse areas of theoretical physics. The possibility of finding an increasing number of 2D surfaces immersed in Lie algebras can be discovered after analysis of the group properties of the models. The main advantage of such an approach appears when group analysis makes it possible to construct algorithms for finding particular classes of 2D surfaces without referring to any additional considerations but proceeding instead directly from the given model under consideration. A systematic method for constructing surfaces and their continuous deformations under various types of dynamics has been extensively developed by many authors (see e.g. \cite{Cies1997, CGS1995,  FG, FGFL, Kono1996}). A broad review of recent developments in this subject can be found in several books \cite{AblClabook, AblSegbook, BobEitbook, CalDegbook, FadTakbook, RogSchbook,  NMPZbook} and references therein.

The methodological approach assumed in this work is based on the Fokas-Gel'fand approach, as presented in \cite{GrundPost2011}, providing a symmetry characterization of deformations of soliton surfaces using the formalism of generalized vector fields and their prolongation structure. The identification of necessary and sufficient conditions for the existence of 2D surfaces in terms of invariance conditions for generalized symmetries allows us to integrate the immersion functions explicitly in terms of conformal transformations in the spectral parameter of the linear spectral problem (LSP), gauge transformation of the wave functions and generalized symmetries of the model and its linear spectral problem. The results obtained for systems of PDE's were so promising that it  is worthwhile to try to adapt this method and check it effectiveness for the case of ordinary differential equations (ODE's) associated with elliptic functions. This is, in short, the aim of the present paper. 

 The crux of the matter is that we consider an ODE in the dependent function $u$ and the independent variable $x$ which can be written in the matrix Lax representation \cite{Lax1968,ZakSha1974, ZakSha1978} involving the differentiation of the Lax pair  with respect to $x$ only. This type of Lax pair of equations is obtained by a compatibility condition of some LSP for which an auxiliary variable, say $y$, has been introduced in the wave function $\Phi$. In this setting, the wave function becomes a function of $\lambda, x, y, u$ and the derivatives of $u$ with respect to $x$. The main advantage of this procedure is that, in using the invariance criterion for the generalized symmetries in terms of their prolongation structure, it leads to simple formulae and allows us to write the explicit form of the soliton surfaces. 

The plan of the paper is as follows. \Sref{prelim} contains a brief account of basic definitions and properties concerning the Fokas-Gel'fand formula for immersion functions in Lie algebras. We concentrate on the study of soliton surfaces associated with ODE's in Lax representation for elliptic functions. In \sref{seclax}, we investigate in detail the Lax pair and find the general solutions for its wave functions for these types of second-order ODE's. These results are then used in \sref{surfaces} to formulate soliton surfaces associated with conformal symmetry in the spectral parameter, gauge transformations of the wave function, and generalized symmetries of the associated integrable ODE and its LSP. In \sref{jac} and \sref{weier}, we present examples of applications of our approach to the case of the Jacobian and $\mathcal{P}$-Weierstrass elliptic functions respectively and calculate their geometric characteristics. \Sref{final} contains final remarks concerning soliton surfaces associated with the elliptic functions, identifies some open questions on the subject and proposes some future developments.

\section{Application of the Fokas-Gel'fand formula to ODE's in Lax representation}\label{prelim}
In this paper, we construct soliton surfaces immersed in Lie algebras using the Fokas-Gel'fand formula for immersion \cite{FGFL}, as formulated in \cite{GrundPost2011}, applied to ODE's. For this purpose, consider an ODE
\be \label{delta} \Delta[u]\equiv \Delta(x, u_x, u_{xx}, \ldots)=0\ee which   admits a Lax pair with potential matrices  $L(\lambda,[u]), \ M(\lambda,[u])$ taking values in a Lie algebra $\g$ which satisfy
\be \label{cc} D_x M+[M,L]=0, \mbox{ whenever } \Delta[u]=0.\ee
Here, $\lambda$ is the spectral parameter. In what follows, we make use of the prolongation structure of vector fields as presented in the book by P. J. Olver \cite{Olver}. For derivatives of $u$ we use the standard notation
\[ \px u=u_x, \qquad \px u_{J}=u_{J, x}, \qquad J =(x,...,x)\]
and, for functions depending on the independent variable $x,$ dependent variable $u$ and its derivatives, the following notation has been used
\[ f[u]=f(x,u_x, u_{xx}, \ldots),\qquad f(\lambda, [u])=f(\lambda, x,u_x, u_{xx}, \ldots).\]
The total derivative in the direction of $x$ takes the form
\be D_x=\px+u_x\frac{\partial }{\partial u} +u_{xx} \frac{\partial }{\partial u_x}+\ldots .\ee

This Lax pair equation \eref{cc} can be regarded as the compatibility conditions of a linear spectral problem (LSP) for wave functions $\Phi$ taking values in the Lie group $G$ with independent variables  $x$ and an auxiliary variable $y,$ and spectral parameter $\lambda.$ For the purpose of symmetry analysis, we allow $\Phi$ to also depend on the dependent variable $u$ as well as its derivatives with respect to $x$. The LSP is written 
\be\label{lsp} \fl D_x\Phi(\lambda, y, [u]) = L(\lambda, [u])\Phi(\lambda, y, [u]), \qquad D_y \Phi(\lambda, y, [u]) =M(\lambda, [u])\Phi(\lambda, y, [u]).\ee 
Here, since the function $u$ is independent of $y$, the total derivative in the direction $y$ is given simply by 
\[ D_y=\frac{\partial}{\partial y}.\]
Note that, because $L$ and $M$ also do not depend on the auxiliary variable $y$, the compatibility conditions for \eref{lsp} are of Lax form \eref{cc}. 

With such an LSP, we can then apply the results of \cite{GrundPost2011}, to construct soliton surfaces immersed in the Lie algebra $\g.$ That is, suppose that $ L, \ M \in \g$ and $\Phi \in G$ satisfy the LSP \eref{lsp} and its compatibility condition \eref{cc}, then there exists a $\g$-valued function $F$ with tangent vectors given by \cite{FGFL}
\be \label{tanF} \px F= \Phi^{-1} A \Phi, \qquad \py F=\Phi^{-1} B \Phi\ee
for any $\g$-valued functions  $A(\lambda, y, [u])$ and $B(\lambda, y, [u])$
which satisfy 
\be \label{AB}  \py A-\px B+[A,M]+[L, B]=0.\ee
Whenever $ A $ and $B $ are linearly independent,  $F$ is an immersion function for a 2D surface in the Lie algebra $\g$. As proved in \cite{GrundPost2011}, three linearly independent terms which satisfy \eref{AB} are 
\bea \label{A} A&=a \frac{\partial }{\partial \lambda} L+\px S +[S,L]+\prw L \in \g,\\
\label{B} B&=a \frac{\partial }{\partial \lambda}  M+\py S +[S,M]+\prw M \in\g,
\eea
where $a=a(\lambda)\in \mathbb{C}$, $S$ is an arbitrary $\g$-valued function of $\lambda, y, x$ as well as the function $u$ and its derivatives,  and $\vec{v}_Q$ is a generalized symmetry of \eref{delta}.  
Further, the $\g$-valued function $F$ can be explicitly integrated as 
\be F=a\Phi^{-1}\frac{\partial \Phi}{\partial \lambda} +\Phi^{-1} S\Phi +\Phi^{-1} \prw \Phi, \ee
as long as  $\vec{v}_Q$ is a generalized symmetry of the linear spectral problem, \eref{lsp}, as well as of the ODE \eref{delta}.

In the next section, we consider a specific form for the ODE which includes differential equations for elliptic functions. We then present explicitly the Lax pair and find its  wave function for the associated LSP.

\section{The Lax Pair and its wave function for a second-order ODE}\label{seclax}
Consider a second-order differential equation given by 
\be \label{uxx} u_{xx}=\frac12 f'(u), \qquad f'(u)=\frac{\partial }{\partial u}f(u) \ee
for some function $f'(u).$ It is straightforward to see that \eref{uxx} admits the first integral 
\be \label{ux} (u_x)^2=f(u), \qquad u_x=\epsilon \sqrt{f(u)},\quad \epsilon^2=1, \ee
and its solutions satisfy
\be \int \frac{du}{\epsilon \sqrt{f(u)}}=x+x_0, \qquad \ x_0 \in \mathbb{R}.\ee 
In the case that 
\be \frac{1}{\sqrt{f(u)}}=R(u, \sqrt{P(u)})\ee
where $R$ is a rational function of its arguments and $P(u)$ is a polynomial of degree 3 or 4, then the function $u$ which solves \eref{uxx} is the inverse of an elliptic integral \cite{AbrStebook, BriBoubook, ByrdFriedman, Incebook}. In particular, this is the case when $f(u)$ is a polynomial of degree 3 or 4.

In this section, we will construct a Lax pair for \eref{uxx} in terms of matrix functions $L, \ M$ taking values in $\glr$ which satisfy the Lax equation \eref{cc} and find solutions for the wave functions $\Phi \in \Glr $ which are solutions of the LSP \eref{lsp}. 
Let us make an assumption for the form of $M$
\be\label{ansm}  M=\left[ \ba{cc} u_x & m_{12}\\ u+\lambda & -u_x \ea \right], \qquad \det(M)=-g(\lambda),\ee
where $g(\lambda)$ is an arbitrary function of $\lambda.$ In what follows, we call $g(\lambda)$ the discriminate. 
From \eref{ansm},  we obtain
\be \det(M)=-g(\lambda)=-u_x^2-(u+\lambda)m_{12},\ee
and  making use of the first integral \eref{ux}, we find that $m_{12}$ has the form 
\be m_{12}=-\frac{f(u)-g(\lambda)}{u+\lambda}.\ee
Note that $m_{12},$ and hence $M$, are rational functions of $\lambda$. Further, if $f(u)$ is a polynomial in $u$, then $m_{12}$ will be a polynomial in $u$ if and only if \[g(\lambda)=f(-\lambda).\]

Next, we solve for $L$ written in the basis of $\glr$
\be L=\ell_1 e_2+\ell_2 e_2+\ell_3 e_3,\ee
where $\ell_i$ $i=1,2,3$ are some functions of $\lambda, x,  u$ and $u_x.$ The matrices $e_i$ written in this basis are
\be \label{basis} e_1=  \left[\ba{cc} 0 & 0 \\ 1 & 0 \ea \right],\quad e_2=\left[\ba{cc} 0 & 1 \\ 0 & 0 \ea \right],\quad e_3=\left[\ba{cc} 1 & 0 \\ 0 & -1\ea \right].\ee
Applying the total derivative $D_x$ to $m_{12}$, we obtain, 
\be D_x(m_{12})=\left(-\frac{f'(u)}{u+\lambda}+\frac{f(u)-g(\lambda)}{(u+\lambda)^2}\right)u_x.\ee
The Lax pair equation \eref{cc} becomes
\bea\label{lspb} \fl \left[\ba{cc} u_{xx} -\ell_1\frac{f(u)-g(\lambda)}{u+\lambda}-\ell_2(u+\lambda) & \left(-\frac{f'(u)}{u+\lambda}+\frac{f(u)-g(\lambda)}{(u+\lambda)^2}\right)u_x +2\ell_2u_x-2\ell_3\frac{f(u)-g(\lambda)}{u+\lambda}\\
u_x-2\ell_1u_x+2\ell_3(u+\lambda) &
 -u_{xx} +\ell_1\frac{f(u)-g(\lambda)}{u+\lambda}+\ell_2(u+\lambda) \ea \right]\nn=\left[\ba{cc} 0 & 0\\ 0&0 \ea\right] .\eea
With the choice of functions, 
\be \ell_1= \frac12, \quad \ell_2=\frac12\left(\frac{f'(u)}{u+\lambda}-\frac{f(u)-g(\lambda)}{(u+\lambda)^2}\right), \qquad \ell_3=0.\ee
\eref{lspb} becomes 
\bea \left[\ba{cc} u_{xx}-\frac12 f'(u) &0 \\ 0 &-u_{xx} +\frac12 f'(u)\ea \right]= \left[\ba{cc} 0 & 0 \\ 0 & 0\ea\right]
,\eea which is equivalent to the ODE  \eref{uxx}. 
Thus, the matrices $L, \ M$ of the form 
\be  \label{LM}L= \frac12\left[\ba{cc} 0 & \frac{f'(u)}{u+\lambda}-\frac{f(u)-g(\lambda)}{(u+\lambda)^2} \\ 1 & 0 \ea \right],\quad  M=\left[ \ba{cc} u_x &-\frac{f(u)-g(\lambda)}{u+\lambda}\\ u+\lambda & -u_x \ea \right]\ee
satisfy the Lax equation \eref{cc}.

Next, we look for the most general solution of the wave function which satisfies the LSP \eref{lsp}. The components of the wave function $\Phi$ are denoted by 
\be \label{phii} \Phi=\left[ \ba{cc} \Phi_{11} &\Phi_{12}\\ \Phi_{21} & \Phi_{22} \ea \right]\in \Glr, \ee
with solutions of the LSP \eref{lsp}
\bea \label{phipmi}  \Phi_{11}=c_1 \phi_{1+} +c_2\phi_{1-}, \qquad & \Phi_{12}=c_3 \phi_{1+} +c_4\phi_{1-},\quad c_i\in \mathbb{R}\\
     \Phi_{21}=c_1 \phi_{2+} +c_2\phi_{2-}, \qquad & \Phi_{22}=c_3 \phi_{2+} +c_4\phi_{2-},\quad i=1,2,3,4 \eea 
and where
\bea \phi_{1\pm}=\frac{\pm \sqrt{g(\lambda)}+u_x}{\sqrt{u+\lambda}}\Psi_{\pm},\\
\phi_{2\pm}={\sqrt{u+\lambda}}\Psi_{\pm},\\
 \label{phipmf}  \Psi_{\pm} =\exp \left[\pm \sqrt{g(\lambda)}\left( y+\int\frac{dx}{2(u+\lambda)}\right)\right].\eea Here the choice of $\epsilon$ comes from \eref{ux}.
Note that, for the purposes of symmetry analysis, it is sometimes useful to express the the integral appearing in \eref{phipmf} as 
\bea \label{intid}\int\frac{dx}{2(u+\lambda)}=\int \frac{du}{2(u+\lambda)u_x}=
\int\frac{\epsilon du}{2 (u+\lambda)\sqrt{f(u)}}\eea
by invoking \eref{ux}. This integral will be explored further for particular choices of $f(u)$ in the following sections. The derivatives of the functions $\Psi_\pm$ have the simple form
\be \label{psixy} \py \Psi _{\pm}= \pm \sqrt{g(\lambda)}\Psi_{\pm}, \qquad D_x \Psi_{\pm}=\pm \frac{\sqrt{g(\lambda)}}{2(u+\lambda)}\Psi_\pm .\ee
We will now check that the wave function $\Phi$ as defined by (\ref{phipmi}-\ref{phipmf}) satisfies the LSP \eref{lsp}. In terms of the functions $\phi_{\alpha \pm}$, the differential equations \eref{lsp} to be verified are reduced to the following
\bea \label{de1} & D_x(\phi_{1\pm})-\left( \frac{f'(u)}{u+\lambda}-\frac{f(u)-g(\lambda)}{(u+\lambda)^2}\right)\phi_{2\pm}=0,\\
\label{de2}  &D_x(\phi_{2\pm})-\frac12 \phi_{1\pm}=0,\\
\label{de3} &\py \phi_{1\pm}-u_x\phi_{1\pm}+\frac{f(u)-g(\lambda)}{u+\lambda}\phi_{2\pm}=0,\\
\label{de4}  &\py \phi_{2\pm}+u_{x}\phi_{2\pm}-(u+\lambda)\phi_{1\pm}=0.\eea
We verify \eref{de1} by computing the following total derivative
\bea \fl D_{x}(\phi_{1\pm})&=\left( \frac{u_{xx}}{\sqrt{u+\lambda}}-\frac12\frac{(\pm \sqrt{g(\lambda)}+u_x)u_x}{\sqrt{u+\lambda}(u+\lambda)}+\frac{\pm\sqrt{g(\lambda)}(\sqrt{g(\lambda)}\pm u_x)}{2(u+\lambda)\sqrt{u+\lambda}}\right)\Psi_\pm\nn
\fl &=\left(\frac{u_{xx}}{(u+\lambda)}+\frac{-u_{x}^2+g(\lambda)}{(u+\lambda)^2}\right)\sqrt{u+\lambda}\Psi_\pm\nn
\fl &=\left(\frac{f'(u)}{2(u+\lambda)}+\frac{-f(u)+g(\lambda)}{(u+\lambda)^2}\right)\phi_{2\pm}\nonumber.\eea
Thus, we can directly observe that \eref{de1} is satisfied for the set of linearly independent solutions $\{ \phi_{1\pm}, \phi_{2\pm}\}$. 
Next, we verify \eref{de2} by computing the total derivative 
\bea D_{x}(\phi_{2\pm})&=\left( \frac{u_x}{2\sqrt{u+\lambda}}+\sqrt{u+\lambda}\frac{\pm\sqrt{g(\lambda)}}{2(u+\lambda)}\right)\Psi_+\nn
&= \frac{u_x\pm\sqrt{g(\lambda)}}{2\sqrt{u+\lambda}}\Psi_+\nn
&=\frac12\phi_{1\pm}.\nonumber\eea
Thus, \eref{de2} holds.
To verify \eref{de3}, we note, using \eref{psixy}, that the dependence of $\Phi$ on $y$ is straightforward  and so the computation is
\bea \fl\py\phi_{1\pm}-u_x\phi_{1\pm}+\frac{f(u)-g(\lambda)}{u+\lambda}\phi_{2\pm}&=(\pm\sqrt{g(\lambda)}-u_x)\phi_{1\pm}+\frac{f(u)-g(\lambda)}{u+\lambda}\phi_{2\pm}\nn
\fl &=\left(\frac{g(\lambda)-u_x^2}{\sqrt{u+\lambda}}+\frac{f(u)-g(\lambda)}{\sqrt{u+\lambda}}\right)\Psi_+\nn
\fl &=0\nonumber.\eea
Thus, \eref{de3} holds. 
The last equation to verify is  \eref{de4},
\bea\fl  \py \phi_{2\pm}+u_{x}\phi_{2\pm}-(u+\lambda)\phi_{1\pm}= (\pm\sqrt{g(\lambda)}+u_x)\phi_{2\pm}-(u+\lambda)\phi_{1\pm}\nn
 =\left( \pm(\sqrt{g(\lambda)}+u_x)\sqrt{u+\lambda}-\frac{(u+\lambda)(\pm\sqrt{g(\lambda)}+u_x)}{\sqrt{u+\lambda}}\right)\Psi_+\nn
=0.\nonumber \eea
We have thus proven that the set of linearly independent solutions $\{ \phi_{1\pm}, \phi_{2\pm}\}$ satisfies (\ref{de1}-\ref{de4}). 
Finally, we note that the requirement $\det(\Phi)=1$ gives an algebraic constraint on the constants $c_i$, \[2\sqrt{g(\lambda)}(c_2c_3-c_1c_4)=1.\]

In the previous section, we have assumed that the Lie algebra is $\glr$. In particular, we assumed that the dependent variable is a real function of a real variable, $x,$  and the auxiliary variable, $y$, to be real as well. On the other hand, we can  consider the complexification of the independent variable $x$, for example in the case of the $\mathcal{P}$-Weierstrass function treated in \sref{weier} which naturally extends to the complex plane. In this case, we can interpret the auxiliary variable $y$ as the complex conjugate of $x$ with the assumption that $u(x)$ is analytic in $x$. The ODE \eref{delta} along with the requirement that $u$ be analytic gives a set of real PDE's for the real and imaginary parts of $u.$ The above construction of $L$ and $M$ still hold, only now they take values in $sl(2,\mathbb{C}).$ The function $f$ is assumed to be a complex function of $u.$ Here the spectral parameter, $\lambda,$ takes values in $\mathbb{C}.$ 

For the purposes of constructing surfaces generated by the wave function $\Phi$, we assume that $u$ is a real function and $x,y$ and $\lambda$ are also real. We choose the constants in $\Phi$ to be
\[c_1=c_2=\frac{1}{2}, \qquad c_3=-c_4=-\frac1{2\sqrt{g(\lambda)}}\]
so that the wave function $\Phi$ simplifies to 
\be\label{phis} \Phi=\left[ \ba{cc} \frac12(\phi_{1+} +\phi_{1-}) & -\frac1{2\sqrt{g(\lambda)}}(\phi_{1+} -\phi_{1-})\\  \frac12(\phi_{2+} +\phi_{2-}) & -\frac1{2\sqrt{g(\lambda)}}(\phi_{2+} -\phi_{2-})\ea \right], \ee
and $\Phi\in \Glr$ for all values of $\lambda \in \mathbb{R}.$ In particular, if $g(\lambda)<0$, then 
\[ \overline{\phi_{\alpha -}}=\phi_{\alpha +}, \quad \alpha=1,2\]  so that their sum will be real and their difference, divided by $\sqrt{g(\lambda)}$, will also be real. 

\section{The induced surfaces}\label{surfaces}

For analytical descriptions of a 2D surfaces, the matrices $A$ and $B$, given in equations \eref{A} and \eref{B} are assumed to be linearly independent. With the wave function $\Phi$, given by \eref{phii} or \eref{phis}, we can construct a surface $F\in \glr$. We construct separately three cases since the pair of matrices $A$ and $B$ corresponds to three types of symmetries. These are a conformal symmetry in the spectral parameter $\lambda$ (called the Sym-Tafel formula), gauge transformations of the wave function and the generalized symmetries of the ODE \eref{uxx}. These three types of symmetries yield different types surfaces. Let us consider each type of them individually. 

\subsection{Sym-Tafel formula for immersion}

The first term in \eref{A} and \eref{B} corresponds to the Sym-Tafel formula for immersion which is given by \cite{Sym, Tafel}
\be F^{ST}=a(\lambda)\Phi^{-1} \frac{\partial }{\partial \lambda} \Phi,\ee
where  $a(\lambda)$ is  an arbitrary function of $\lambda.$ This surface $F^{ST}$ has tangent vectors of the form 
\be \label{STtan} \px F^{ST}=\Phi^{-1}(\frac{\partial }{\partial \lambda} L)\Phi, \qquad \py F^{ST}=\Phi^{-1}(\frac{\partial }{\partial \lambda} M) \Phi.\ee
With $L$ and $M$ given by \eref{LM}, the tangent vectors in \eref{STtan} are linearly independent and so the function $F^{ST}$ gives an immersion of a 2D surface in the Lie algebra $\glr.$

\subsection{Surfaces associated with gauge symmetry}

The second term in \eref{A} and \eref{B} corresponds to the gauge symmetry of the linear spectral problem \eref{lsp}. The surface $F^S$ corresponding to this gauge term can be integrated explicitly as \cite{ Cies1997, FG, FGFL}
\be F^S=\Phi^{-1} S(\lambda, y, [u]) \Phi\ee
with tangent vectors 
\be D_x F^S=\Phi^{-1}\left( D_x S+[S,L]\right)\Phi, \qquad   \py F^S=\Phi^{-1}\left( \py S+[S,M]\right)\Phi.\ee
Again, for $F$ to be an immersion, we require the linear independence of the tangent vectors.

Note that, for any surface $P\in\glr$, $P$ can be expressed as 
\be \label{P} P=\Phi^{-1}S\Phi=F^{S} \ee 
and hence $F^{S}$ represents a completely arbitrary surface immersed in the Lie algebra. 
For any $\glr$-valued function $S,$ written in the basis \eref{basis} as
\be S=s_1(\lambda,y,[u])e_1+s_2(\lambda,y,[u])e_2 +s_3(\lambda,y,[u]) e_3, \ee 
the surface $F^S$ takes the form 
\bea \fl F^{S}&=s_1\left[\frac{\left((u_x+\sqrt{g})\Psi_+-(u_x-\sqrt{g})\Psi_-\right)^2}{4(u+\lambda)}(e_1-e_2)+ \frac{(u_x+\sqrt{g})^2\Psi_+^2-(u_x-\sqrt{g})^2\Psi_-^2}{(u+\lambda)\sqrt{g}}e_3\right]\nn
\fl &+s_2\frac{(u+\lambda)}{4}\left(-(\Psi_++\Psi_-)^2e_1+\frac{1}{g}(\Psi_+-\Psi_-)^2e_2-\frac{1}{\sqrt{g}}(\Psi_+^2-\Psi_-^2)e_3\right)\nn
\fl & + s_3\Bigg[\frac12(\Psi_++\Psi_-)\left((u_x+\sqrt{g})\Psi_++(u_x-\sqrt{g})\Psi_-\right)e_1\nn
\fl \label{SC}&+\frac{(\Psi_+-\Psi_-)\left((u_x+\sqrt{g})\Psi_+-(u_x-\sqrt{g})\Psi_-\right)}{2g}e_2 -\frac{(u_x+\sqrt{g})\Psi_+^2-(u_x+\sqrt{g})\Psi_-^2}{2\sqrt{g}}e_3\Bigg]\nonumber.\eea
We can interpret the surface given in the form \eref{P} as an arbitrary surface immersed in this Lie algebra written in the frame defined by conjugation by the wave function $\Phi,$ an element of the Lie group $\Glr$.

\subsection{Surfaces associated with generalized symmetries}

The third term of \eref{A} and \eref{B} is associated with generalized symmetries of equation \eref{we}. That is, suppose that there exists a generalized vector field with evolutionary representative
\be \label{vq} \vec{v}_Q=Q[u]\frac{\partial }{\partial u}\ee
which is a generalized symmetry of \eref{uxx} in the sense that 
\be \prw \left(u_{xx}-\frac12 f'(u)\right)=0, \qquad \mbox{ whenever } u_{xx}-\frac12 f'(u)=0\ee
holds, where we have used the standard definition of the prolongation of a generalized vector field given as in the book by P. J. Olver \cite{Olver},
\be pr\vec{v}_Q=Q\frac{\partial }{\partial u}+D_xQ\frac{\partial }{\partial u_x}+D_x^2Q\frac{\partial }{\partial u_{xx}}+\ldots.\ee
The determining equations for $Q$ are thus
\be \label{detQ} D_x^2Q-\frac12f''(u)Q=0, \quad  \mbox{ whenever } u_{xx}-\frac12 f'(u)=0.\ee
For such a generalized symmetry \eref{vq}, there exists a surface immersed in the Lie algebra $\g$, say $F^Q$, with tangent vectors \cite{GrundPost2011}
\be \label{Fq} D_x F^Q=\Phi^{-1} \prw L \Phi, \qquad \py F^Q=\Phi^{-1} \prw M \Phi.\ee
Further, if the generalized symmetry is also a symmetry of the LSP in the sense that 
\bea \prw\left(\px \Phi-L\Phi\right)=0, \qquad \mbox{ whenever } \px \Phi-L\Phi=0\\
 \prw\left(\py \Phi-M\Phi\right)=0, \qquad \mbox{ whenever } \py \Phi-M\Phi=0,\eea
then  \eref{Fq} can be integrated and the surface $F^Q$ is given, up to a constant matrix of integration in $\glr$ by \cite{GrundPost2011}
\be \label{FQ} F^Q=\Phi^{-1} \prw \Phi.\ee

The following characteristics $Q_i$'s are solutions of the determining equation \eref{detQ}
\begin{enumerate}
\item $Q_1\equiv u_x$
\item $Q_2\equiv u_x\int f(u)^{-\frac32}{du}$
\item $Q_3\equiv xu_x+\gamma u$, this is only in the special case $f(u)=c_1+c_2u^{\ell}$ for $ \ell=2(1+1/\gamma),$ $\gamma, c_1, c_2\in \mathbb{R}. $
\end{enumerate}

We begin our consideration with $Q_1=u_x.$ It can be observed directly that the differential equation \eref{uxx} is invariant under translation in $x.$ Thus, the vector field 
\be \vec{v}_{u_x}=u_x\frac{\partial}{\partial u} \ee
is a symmetry of \eref{uxx}. In fact, for any $G(y, \lambda, [u])$ which does not explicitly depend on $x$, the prolongation of $\vec{v}_{u_x}$ acts as a total derivative, 
\be pr\vec{v}_{u_x}(G(y,\lambda,[u])=D_x(G(y,\lambda,[u])) \iff \frac{\partial}{\partial x}G(y,\lambda,[u])=0.\ee
Hence, we can see that 
\be \fl  pr \vec{v}_{u_x}(u_{xx}-\frac12f'(u))=D_x\left(u_{xx}-\frac12f'(u)\right)=0,\qquad  \mbox{ whenever } u_{xx}-\frac12f'(u)=0.\ee
Similarly, the wave functions and the potential matrices $L$ and $M$ do not depend explicitly on $x$ and so 
\bea \fl pr\vec{v}_{u_x}\left(D_x\Phi -L\Phi\right)=D_x\left(D_x\Phi -L\Phi\right)=0,\qquad  \mbox{ whenever } D_x\Phi -L\Phi=0,\\
    \fl pr\vec{v}_{u_x}\left(D_{y} \Phi -M\Phi\right)=D_x\left(D_{y} \Phi -M\Phi\right)=0, \qquad   \mbox{ whenever } D_{y} \Phi -M\Phi=0.\eea
Thus, for an arbitrary function $f(u)$, the vector field $\vec{v}_{u_x}$ is a symmetry of both \eref{uxx} and the LSP \eref{lsp} and so according to \cite{GrundPost2011}, there exists a surface defined by the immersion function 
\be F^{u_x}=\Phi^{-1} pr \vec{v}_{u_x}\Phi=\Phi^{-1}D_x\Phi\ee
with tangent vectors 
\be D_x F^{u_x}=\Phi^{-1} D_x(L) \Phi, \qquad \py F^{u_x}=\Phi^{-1} D_x(M) \Phi.\ee

For the second case, we can verify that  $Q_2=u_x\int{f(u)^{-\frac32}}du$ solves the determining equation \eref{detQ}. The action of $pr\vec{v}_{Q_2}$ on the LSP is given by 
\bea\fl  pr\vec{v}_{Q_2}(\py \Phi-M\Phi)=\frac{u_x}{\sqrt{u+\lambda}\sqrt{f(u)}}\left[\ba{cc}-(\Psi_++\Psi_-), & g(\lambda)^{-\frac12}(\Psi_+-\Psi_-)\\ 0,&(\Psi_++\Psi_-)\ea\right],\\
\fl pr\vec{v}_{Q_2}(D_x \Phi-L\Phi)=\frac{u_x}{2(u+\lambda)^{\frac32}\sqrt{f(u)}}\left[\ba{cc}-(\Psi_++\Psi_-) & g(\lambda)^{-\frac12}(\Psi_+-\Psi_-)\\ 0&(\Psi_++\Psi_-)\ea\right].
\eea
Since these quantities do not vanish for all solutions of the LSP, the vector field $\vec{v}_{Q_2}$ is not a generalized symmetry of the LSP. 
Thus, while there exists an $\glr$-valued immersion function $F^{Q_2}$ with tangent vectors 
\be D_x F^{Q_2}=\Phi^{-1} pr\vec{v}_{Q_2}( L) \Phi, \qquad \py F^{Q_2}=\Phi^{-1} pr\vec{v}_{Q_2}(M) \Phi,\ee
the immersion function $F^{Q_2}$ is not of the form given in \eref{FQ}.

Similarly, for the special case $Q_3\equiv xu_x+\gamma u$ when $f(u)=c_1+c_2u^{\ell},$ it is straightforward to verify that $Q_3$ satisfies the determining equation \eref{detQ}. However, its action on the LSP gives 
 \bea \fl pr\vec{v}_{Q_3}(\py \Phi-M\Phi)=\frac{c_1(1+\gamma)}{\sqrt{u+\lambda}}\left[\ba{cc}-(\Psi_++\Psi_-) & g(\lambda)^{-\frac12}(\Psi_+-\Psi_-)\\ 0&(\Psi_++\Psi_-)\ea\right]\\
\fl pr\vec{v}_{Q_3}(D_x \Phi-L\Phi)=\frac{c_1(1+\gamma)}{2(u+\lambda)^{\frac32}}\left[\ba{cc}-(\Psi_++\Psi_-) & g(\lambda)^{-\frac12}(\Psi_+-\Psi_-)\\ 0&(\Psi_++\Psi_-)\ea\right].\eea
Thus, unless $c_1=0$ or $\gamma=-1$, $\vec{v}_{Q_3}$ is not a symmetry of the LSP. In the former case, the differential equation reduces to the degenerate case $ u_{x}^2=c_2u^{\ell}$ and in the latter case $u$ is linear in $x$ since $ u_x^2={c_1+c_2}.$

Thus, the surface associated with generalized vector field $\vec{v}_{Q_1}$ can be integrated explicitly and is given by \eref{FQ} whereas the surfaces associated with the $\vec{v}_{Q_2}$ and $\vec{v}_{Q_3}$ are only of the form \eref{FQ} for special cases. However, for all generalized symmetries there exists a surface with tangent vectors given by \eref{Fq} and we can study their geometric properties based on the tangent vectors to the surface. To this end, in the next section we give a scalar product on the tangent spaces for the surfaces.

\subsection{Induced metrics on the surfaces}
Here, we would like to introduce two possible choices for an induced metric on the tangents to the surface $F\in \glr.$ We take the basis 
for $\glr$ given by \eref{basis}. 
A first choice for a metric would be to decompose the matrix in the basis $\{ e_1, e_2, e_3\}$ and then to use the standard Euclidean metric. That is, given $X, Y\in \glr$ with 
\be X=X^ie_i, \qquad  Y=Y^ie_i, \qquad i=1,2,3\ee then the inner product and norm in Euclidean space is defined by 
\be \langle X, Y \rangle=X^iY^i, \qquad || X||=\sqrt{X^iX^i}.\ee
This constitutes an inner product on the tangent vectors for the  surface $F\in \glr.$ 

On the other hand, we can also use a symmetric bilinear product defined using the Killing form, $B(X,Y).$  The main advantage of this form is that it is invariant under conjugation by the group and so, because of the form of the tangent vectors \eref{tanF}, the geometric quantities associated with the surfaces will be independent of the wave function $\Phi \in \Glr.$
The Killing form on $\glr$ is given, up to a normalization factor,  by (see e.g. \cite{Helgason}) 
\be B(X,Y)=\tr(XY).\ee
In terms of the basis $\{ e_1, e_2, e_3\}$, the matrices $X,Y\in \glr $ and the Killing form $B(X,Y)$ can be represented as the following
\be X=\left[\ba{c} X^1 \\ X^2\\X^3 \ea \right], \ Y=\left[\ba{c} Y^1 \\ Y^2\\Y^3 \ea \right], \qquad  B(X,Y)=X^T B Y, \ee 
with
\be B(X,Y)=\left[ \ba{ccc} 0 & 1 & 0\\ 1 & 0 &0\\ 0& 0&2\ea \right] .\ee
The Killing form has signature $(2,1)$ and so induces a pseudo-Euclidean metric on the tangents to the surface given by the immersion function, $F\in \glr.$ 
The surface defined via the immersion function $F$ with the Euclidean metric $\langle \ , \ \rangle$ is a Riemannian manifold while the surface with the Killing form $B$ is a pseudo-Riemannian manifold \cite{DoCarmo, Willmorebook}.

As an example, for a general $f(u)$ the first fundamental form for the surfaces $F^{ST}$ and $F^{u_x}$  with the pseudo-Euclidean metric are given by  
\be I_B(F^{ST})=\left(2\frac{f-g}{(u+\lambda)^3}-\frac{f'-g'}{(u+\lambda)^2}\right)dxdy+2\left(\frac{g'}{v+\lambda}+\frac{f-g}{(u+\lambda)^2}\right)dy^2
,\ee
and 
\be\fl  I_B(F^{u_x})=\left(\frac{ff''}{u+\lambda}-\frac{2ff'}{(u+\lambda)^2}+\frac{2f(f-g)}{(u+\lambda)^3}\right)dxdy+\left(\frac{f'^2}{2}-\frac{2ff'}{u+\lambda}+\frac{2f(f-g)}{(u+\lambda)^2}\right)dy^2,\ee
where for convenience, we denote 
\[ f=f(u), \quad g=g(\lambda), \qquad f'=\frac{\partial}{\partial u}{f(u)}, \quad g'=\frac{\partial}{\partial \lambda }{g(\lambda)}.\]
In the Euclidean metric, the components of the metrics are significantly more complicated and can be found in \ref{appendixeuc}. Similarly, the components of the metric for $F^{S}$, in both the Euclidean and Pseudo-Euclidean metric, are too involved to write out in an illustrative fashion. However, they are directly computable from \eref{SC}. In practice, the gauge term can be used to simplify the expressions for the surfaces. 

\section{Jacobian elliptic functions and associated surfaces}\label{jac}
Consider the differential equation for the Jacobian Elliptic functions
\be \label{xnx} (u_x)^2=(1-u^2)(k_1+k_2u^2),\ee
or alternatively
\be \label{xn} u_{xx}=-2k_2u^3+(k_2-k_1)u.\ee
Solutions of \eref{xnx} are given by Jacobian elliptic functions, for the choice of constants given in \tref{Jaccon} with 
\be k'^2+k=1, \qquad 0\leq k, k' \leq1, \ee
\begin{table}
\caption{\label{Jaccon}Constants in \eref{xnx} for the Jacobian Elliptic functions.}
\begin{indented}
\item[]\begin{tabular}{@{}rrr}
\br
$k_1$&$ k_2$& Solutions of \eref{xnx}\\
\mr
$1$&$ -k^2 $&$sn(x, k)$\\ 
$k'^2$&$ k^2$ & $cn(x,k)$\\
$-k'^2$&$ 1$ & $dn(x,k)$\\
\br
\end{tabular}
\end{indented}
\end{table}
In terms of the notation of \sref{seclax}, the function $f(u)$ takes the form \be f(u)=(1-u^2)(k_1+k_2u^2).\ee 
Choosing 
\be g(\lambda)=f(-\lambda)=(1-\lambda^2)(k_1+k_2\lambda^2),\ee the matrices $L$ and $M$ become 
\bea M=\left[\ba{cc} u_x& (u-\lambda)(k_2(u^2+\lambda^2)+k_1-k_2)\\ u+\lambda& -u_x\ea \right],\\
L=\frac12\left[\ba{cc} 0&  -3k_2u^2+2\lambda k_2u +k_1-k_2-k_2\lambda^2 \\1 &0\ea\right].\eea 
Taking the normalization of $\Phi$ as in \eref{phis}, the wave function takes the form 
\begin{equation}\label{Phif} \fl \Phi=\left[\ba{cc} \frac{(\sqrt{g(\lambda)}-u_x)\Psi_+-(\sqrt{g(\lambda)}+u_x)\Psi_-}{2\sqrt{u+\lambda}},& \frac{(\sqrt{g(\lambda)}+u_x)\Psi_--(\sqrt{g(\lambda)}-u_x)\Psi_+}{2\sqrt{g(\lambda)}\sqrt{u+\lambda}}\\ \frac{\sqrt{u+\lambda}(\Psi_++\Psi_-)}{2}, & \frac{\sqrt{u+\lambda}(\Psi_--\Psi_+)}{2\sqrt{g(\lambda)}} \ea\right]\nonumber,\end{equation}
where 
\be \Psi_\pm=\exp\left[\sqrt{g(\lambda)}\left(y+\int \frac{dx}{2(u+\lambda)}\right)\right].\ee
Recall that, for the purpose of computing the action of the generalized vector field $\vec{v}_Q$ on $\Phi,$ it is convenient to use the identity \eref{intid}
\be \label{intjac}  \int \frac{dx}{2(u+\lambda)}=\int\frac{\epsilon du}{2 (u+\lambda)\sqrt{(1-u^2)(k_1+k_2u^2)}}, \ee when $u$ is a solution of \eref{xnx}.
The integrated form of \eref{intjac}  is
\bea\fl  \int \frac{\epsilon du}{2 (u+\lambda)\sqrt{(1-u^2)(k_1+k_2u^2)}}=\frac{\epsilon}{2\lambda\sqrt{k_1}}\Pi\left(u,\frac{1}{\lambda^2}, \sqrt{\frac{-k_2}{k_1}}\right)\nn
-\frac{\epsilon}{4\sqrt{g(\lambda)}}tanh^{-1}
\left(\frac{(k_2-k_2-2k_2\lambda^2)u^2+(k_2-k_1)\lambda^2+2k_1}{2\sqrt{g(\lambda)}\sqrt{(1-u^2)(k_1+k_2u^2)}}\right)+c_0,\eea
where $\Pi\left(u,a, b\right)$ is the normal elliptic integral of the third kind, see e.g. \cite{ByrdFriedman} 
\be\label{Pi} \Pi\left(u,\alpha^2,k \right)=\int_0^x\frac{dt}{(1-\alpha^2t^2)\sqrt{1-t^2}\sqrt{1-k^2t^2}}.\ee
The functions $\Psi_\pm$ are then given by 
\bea \fl \Psi_\pm &=\exp\left[\pm\sqrt{g(\lambda)}\left(y+ \frac{\epsilon}{\lambda\sqrt{k_1}}\Pi\left(u,\frac{1}{\lambda^2}, \sqrt{\frac{-k_2}{k_1}}\right)+c_0\right)\right]\nn
\fl &\times\left[\frac{2\sqrt{g(\lambda)}\sqrt{(1-u^2)(k_1+k_2u^2)}+(k_2-k_2-2k_2\lambda^2)u^2+(k_2-k_1)\lambda^2+2k_1}{2\sqrt{g(\lambda)}\sqrt{(1-u^2)(k_1+k_2u^2)}-(k_2-k_2-2k_2\lambda^2)u^2-(k_2-k_1)\lambda^2-2k_1}\right]^{\mp \frac{\epsilon}4}.\eea
In the graphs below, we choose the integration constant $c_0$ so that $\Psi_\pm(0,0)=1.$

%\left[\frac{\sqrt{1-\lambda^2}dn(x,k)-\sqrt{k^2\lambda^2+k'^2}sn(x,k)}{\sqrt{1-\lambda^2}dn(x,k)+\sqrt{k^2\lambda^2+k'^2}sn(x,k)}\right]^{\pm\frac14}.\eea

From the wave function $\Phi$, it is immediate to compute the analytical form of the surface generated by the terms described in \sref{surfaces},
\bea F=\left(a(\lambda)\Phi^{-1}\frac{\partial}{\partial \lambda} \Phi+\Phi^{-1}S \Phi+b \Phi^{-1}D_x\Phi\right)\in \glr \qquad a(\lambda), b\in \mathbb{R}\nonumber.\eea
Below, we have presented several graphs of the induced surfaces plotted with the help of MAPLE. The surfaces in figure 1 have constants chosen so that the discriminant is negative, $g(\lambda)<0,$ and so the surfaces behave like trigonometric functions whereas the surfaces in figure 2 have positive discriminant, $g(\lambda)>0,$ and so have exponential type behavior. 
\begin{figure}\label{gneg}
\begin{center}$
\begin{array}{ccc}
\includegraphics[width=2.0in]{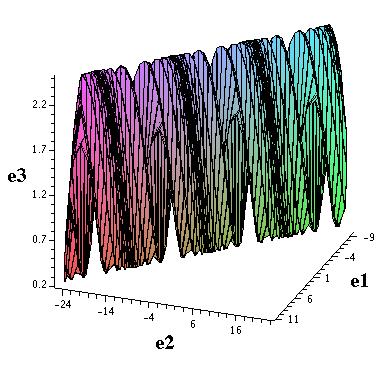} &
\includegraphics[width=2.0in]{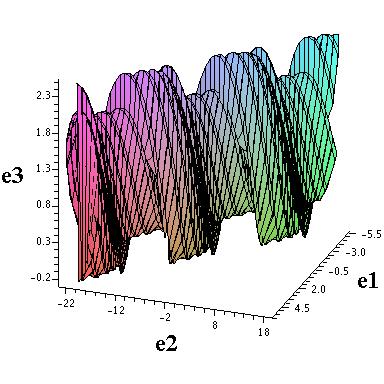}&
\includegraphics[width=2.0in]{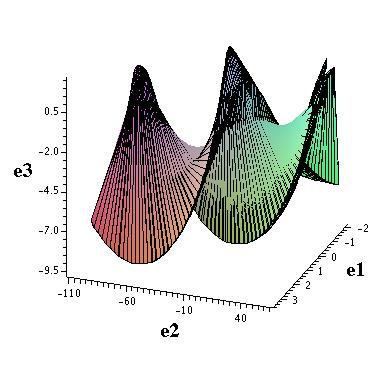} \\
F^{ST}: \lambda=1.2, \ k=0, & F^{ST}: \ \lambda=1.2, \ k=0.5,&F^{ST}: \ \lambda=1.2, \ k=0.8,\\
\includegraphics[width=2.0in]{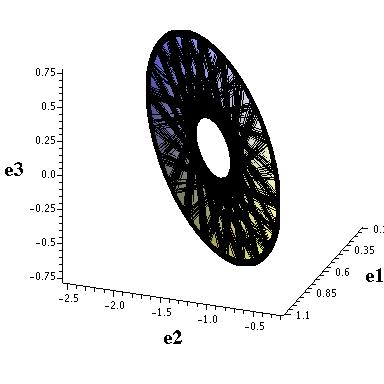} &
\includegraphics[width=2.0in]{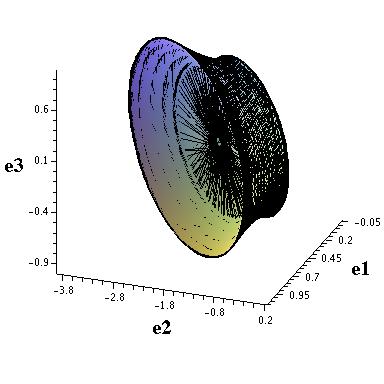}&
\includegraphics[width=2.0in]{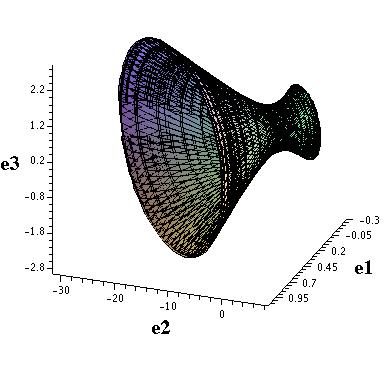} \\
F^{u_x}: \lambda=1.2, \ k=0, & F^{u_x}: \ \lambda=1.2, \ k=0.5,&F^{u_x}: \ \lambda=1.2, \ k=0.8, \end{array}$

\end{center}
\caption{Surfaces $F^{ST}$ and $F^{u_x}$ for $u=sn(x,k)$ with the discriminant $g(\lambda)<0$ and  $x$ and $y \in [-8,8]$. The axes indicate the components of the immersion function in the basis \eref{basis}  }
\end{figure}

\begin{figure}\label{gpos}
\begin{center}$
\begin{array}{cc}
\vspace{-10pt}
\includegraphics[width=2.5in]{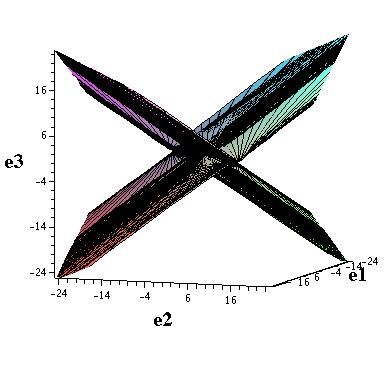} &
\includegraphics[width=2.5in]{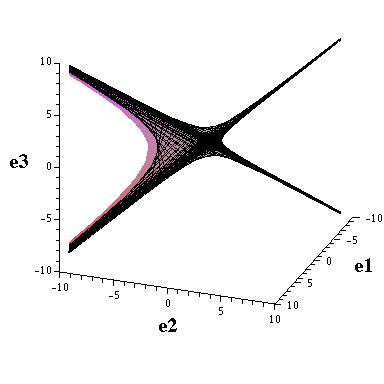}\\
F^{ST}: \lambda=0.5, \ k=0.2 & F^{u_x}: \lambda=0.5, \ k=0.2\end{array}$
\end{center}
\caption{Surfaces  $F^{ST}$ and $F^{u_x}$ for $u=sn(x,k)$ with the discriminant $g(\lambda)>0$,  $x\in [-20,20]$ and $y \in [-5,5]$. The axes indicate the components of the immersion function in the basis \eref{basis} }
\end{figure}

In the metric induced by the Killing form, the first fundamental form for the surface $F^{ST}$ is 
\[ \fl  I_{B}(F^{ST}) =\frac{a(\lambda)^2}2(k^2(u-\lambda)dxdy-\frac{a(\lambda)^2}2\left[k^2(u^2-2\lambda u+3\lambda^2-2)+1\right]dy^2.\]
 and for $F^{u_x}$ is 
\bea \fl I_{B}(F^{u_x})=-\frac{k^2b^2}{2}(3u-\lambda)(1-u^2)(k^2u^2+k'^2)dxdy\nn
\fl \qquad +\frac{b^2}2\bigg[{k}^{4}{u}^{6}+2\,{u}^{5}{k}^{4}\lambda-{u}^{4}{k}^{4}{\lambda}^{2} 
 -\left(4{k}^{4}\lambda-2\lambda{k}^{2}\right){u}^{3}+\left(3{k}^{2}+2{k}^{4}{\lambda}^{2}-3{k}^{4}-{k}^{2}{\lambda}
^{2}\right){u}^{2}
\nn\fl\qquad 
 -\left(2\lambda{k}^{2}-2{k}^{4}\lambda
 \right) u-{k}^{4}{\lambda}^{2}+2{k}^{4}+{k}^{2}{\lambda}^{2}-3{k}
^{2}+1\bigg]dy^2\nonumber.\eea
We can also study the surfaces associated with generalized symmetries $Q_2= u_x\int f(u)^{-\frac32}du$ and 
 $Q_3= xu_x+\gamma u$, in the special case when $k_1=-k_2,\ \gamma=1 $ and $f(u)=k_1(1-u^{4}),$ but the immersion functions are not of the form \eref{FQ} since the vector fields $\vec{v}_{Q_2}$ and $\vec{v}_{Q_3}$ are not generalized symmetries of the LSP. However, we can study the geometric properties of these surfaces via their tangent vectors.

\section{$\mathcal{P}$-Weierstrass elliptic function and associated surfaces}\label{weier}
Consider the differential equation for the $\mathcal{P}$-Weierstrass
 elliptic function 
\be u_x=-\sqrt{4u^3-{g_2}u-g_3},\ee
or alternatively
\be \label{we} u_{xx}-6u^2+\frac{g_2}2=0.\ee
In terms of the notation of \sref{seclax}, \be f(u)=4u^3-g_2u-g_3.\ee 
If we choose
\be g(\lambda)=f(-\lambda),\ee the Lax pair $L, \ M$ becomes 
\bea \label{72} M=\left[\ba{cc} u_x& -4u^2+4\lambda u-4\lambda^2 +g_2\\ u+\lambda& -u_x\ea \right],\\
\label{73} L=\frac12\left[\ba{cc} 0&  4u-2\lambda\\1 &0\ea\right].\eea 

Taking the normalization of $\Phi$ as in \eref{phis}, the wave function takes the form given by \eref{Phif} but differs from the wave functions from the previous section since the functions $\Psi_\pm$ are instead given by 
\be \Psi_\pm=\exp\left[\sqrt{g(\lambda)}\left(y+\int \frac{dx}{2(u+\lambda)}\right)\right].\ee
Again, for computing the action of the generalized vector $\vec{v}_Q$ field on $\Phi,$ it is convenient to use the identity
\be \label{intp} \int \frac{dx}{2(u+\lambda)}=-\int\frac{ du}{2 (u+\lambda)\sqrt{4u^3-g_2u+g_3}}, \ee when $u$ is a solution of \eref{we}.
The integrated form of \eref{intp} is
\bea\fl\int -\frac{du}{2 (u+\lambda)\sqrt{4u^3-g_2u-g_3}}=\nn
\frac{-1}{2(\lambda-a_1)(2a_1+a_2)}\Pi\left(\sqrt{\frac{u+a_1}{a_1-a_2}}, \frac{a_1-a_2}{\lambda-a_1},\sqrt{ \frac{a_1-a_2}{2a_1+a_2}}\right)+c_0,\eea
where, for simplicity, we have used the constants $a_1, \ a_2$ to represent the roots of the polynomial $f(u)$ as in 
\be f(u)=4u^3-g_2u-g_3=4(u+a_1)(u+a_2)(u-a_1-a_2).\ee
Here, $\Pi$ is the solution of the elliptic integral defined by \eref{Pi}. In the graphs below, we choose the integration constant $c_0$ so that $\Psi_\pm(0.5,0)=1.$

From the wave function, $\Phi$ it is immediate to compute the analytical form of the surface generated by the terms described in \sref{surfaces},
\bea F=\left(a(\lambda)\Phi^{-1}\frac{\partial}{\partial \lambda} \Phi+\Phi^{-1}S \Phi+b \Phi^{-1}D_x\Phi\right)\in \glr \qquad a(\lambda), b\in \mathbb{R}\nonumber.\eea
In figure 3, we have included graphs of the induced surfaces for $\mathcal{P}$-Weierstrass functions with $g_2=0, g_3=1$, i.e. on an equianharmonic lattice \cite{AbrStebook}. Again, the surfaces exhibit periodic behavior when $g(\lambda)<0$ and exponential behavior when $g(\lambda)>0$.  
\begin{figure}\label{weiergraph}
\begin{center}$
\begin{array}{cc}
\vspace{-10pt}
\includegraphics[width=2.5 in]{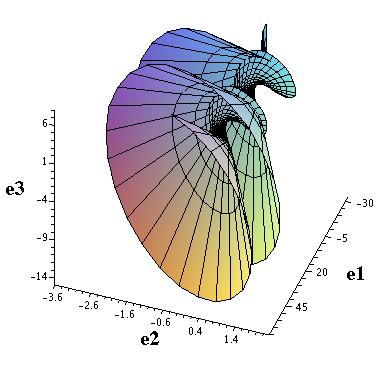} &\includegraphics[width=2.5in]{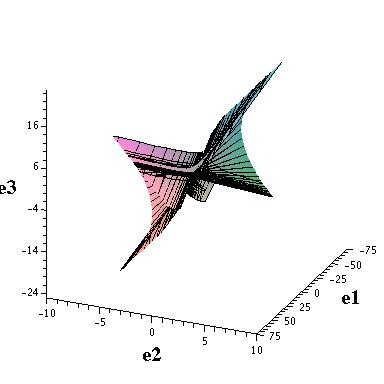} \\
F^{ST}: \lambda=1, \ g(\lambda)=-5 & \\
\includegraphics[width=2.5in]{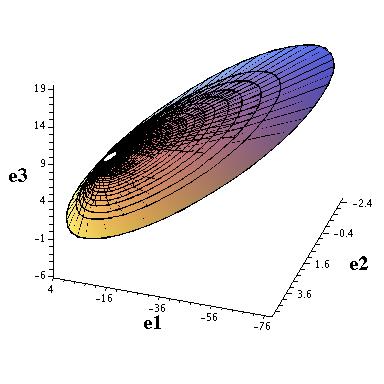}& \includegraphics[width=2.5in]{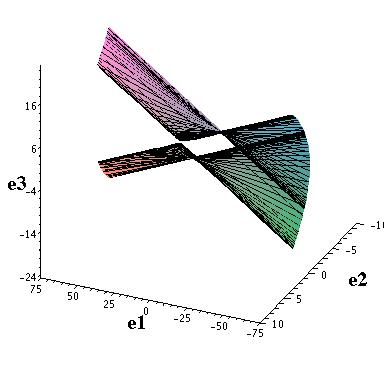} \\
F^{u_x}: \lambda=1, \ g(\lambda)=-5 & F^{u_x}: \lambda=-5, \ g(\lambda)=31  \end{array}$
\end{center}
\caption{Surfaces  $F^{u_x}$ for $u=\mathcal{P}(x,0,1)$ with  $x\in [0.2,3]$ and $y \in [-\pi/g(1),\pi/g(1)]$ for the negative discriminate cases and $x\in [-1,1]$ $y\in [-0.5,0.5]$ for the positive discriminant cases. The axes indicate the components of the immersion function in the basis \eref{basis} }
\end{figure}
We can also study the surfaces associated with generalized symmetries $Q_2= u_x\int f(u)^{-\frac32}du$ and 
 $Q_3= xu_x+\gamma u$, in the special case where $g_2=0,\ \gamma=2 $ and $f(u)=4u^{3}-g_3.$ However, the immersion functions are not of the form \eref{FQ} except in the latter case when $g_3$ is also equal to zero and the solutions of \eref{we} are rational functions of $x.$

In the metric induced by the Killing form, the first fundamental form for the surface $F^{ST}$ is 
\[ I_B(F^{ST})=-2a^2(\lambda)\left(dxdy+3a^2(2\lambda-u)dy^2\right).\]
and, for $F^{u_x},$ is 
\bea  I_B(F^{u_x})&=4b^2\Bigg[(4u^3-g_2u-g_3)dxdy\nn&+\left(2u^4+8\lambda u^3+g_2u^2-2(g_2\lambda-2g_3)u-2\lambda g_3+\frac18g_2^2\right)dy^2\Bigg]\nonumber.\eea
Both fundamental forms depend on the solution $u$ and the spectral parameter $\lambda$. Note that, in both cases, the tangent vectors in the $x$-direction are null vectors in the pseudo-Euclidean metric. 

\section{Final remarks}\label{final}
In this paper, we have discussed certain classes of surfaces immersed in Lie algebras associated with elliptic integrals. This problem has been studied recently for PDE's by the authors in \cite{GrundPost2011} using a symmetry characterization of continuous deformations of soliton surfaces immersed in Lie algebras based on the formalism of generalized vector fields and their prolongation structure. The necessary and sufficient conditions for the existence of such surfaces in terms of the invariance conditions has been established for integrable PDE's. In this context, we have adapted the proposed procedure for integrable ODE's admitting a Lax representation \eref{cc} and shown, as in the PDE case, that the problem requires the examination of conformal symmetries in the spectral parameter, gauge transformations of the wave function and generalized symmetries of the associated model and its LSP. To perform this symmetry analysis, we have constructed a Lax pair for a second-order ODE which includes, among others, the case of elliptic integrals. Next, we solved the LSP and found explicitly the most general form of the wave function. Next, we constructed surfaces in $\glr$ by analytic methods for a general case where the right hand side of \eref{ux} is an arbitrary function $f(u)$. We were able to explicitly integrate these ODE's in terms of elliptic integrals if $f(u)$ is a polynomial in $u$ of degree $3$ or $4$ and give explicit forms of the corresponding soliton surfaces for the Jacobian and $\mathcal{P}$-Weierstrass elliptic functions. A geometrical analysis of these surfaces has been performed by using an appropriate inner product which allows for the construction of  Riemannian and pseudo-Riemannian manifolds. The elaborate procedure was applied to examples and we have given the first fundamental form for the surfaces as well as graphs of the surfaces for a range of parameters leading to diverse types of surfaces.  

Additional questions which could be asked involve the geometric properties of the surfaces associated with elliptic integrals. It would be especially interesting  to determine some global characteristics of the soliton surfaces defined by the immersion function $F$. These quantities, such as the Willmore functional or Euler-Poincar\'e characteristics as well as the Gaussian and mean curvatures, can be calculated but the expressions are rather involved, so we omit them here. It is an open question as to whether, by judicious choice of the linearly independent terms in the matrices $A$ and $B$ given by \eref{A} and \eref{B}, the equations can be simplified. Finally, another interesting avenue for future research could include the application of the methods presented here to other integrable equations ODE's, for example hyperelliptic functions, which admit a Lax representation and to study their soliton surfaces.

\ack The authors would like to thank Professor R Conte (Ecole Normale Sup\'erieure, CMLA, Cachan and Centre d'Energie Atomique de Saclay) for helpful and interesting discussions on the topic of this paper. The research reported in this paper is supported by NSERC of Canada. S Post acknowledges a postdoctoral fellowship provided by the Laboratory of Mathematical Physics of the CRM, Universit\'e de Montr\'eal. 
\appendix 

\section[]{Components of the first fundamental form for surfaces in the Euclidean metric}\label{appendixeuc}
The components of the first fundamental form for the surface induced via the Sym-Tafel formula for immersion, in the Euclidean metric, are given by 
\bea\fl \langle F^{ST}_x, F^{ST}_x\rangle =\left(\frac{(f'-g')(u+\lambda)-2(f-g)}{64(u+\lambda)^2g(\lambda)}\right)^2\nn
\fl \qquad  \times \left((g^2+g+1)(\Psi_+^4+\Psi_-^4)+4(g^2-1)(\Psi_+^2+\Psi_-^2)+2(3g^2-g+3)\right), \nn
\fl \langle F^{ST}_x, F^{ST}_y\rangle =\frac{2(f-g)-(f'-g')(u+\lambda)}{32(u+\lambda)^3g(\lambda)^{\frac32}}\times\Bigg[2(3g^2-g+3)(u+\lambda)g'-4g(g^2+g+1)\nn
\fl\qquad -2\left(\Psi_+^4-1\right)\left((g^2+g+1)(1+\Psi_-^4)+2(g^2-1)\Psi_-^2\right)u_x\nn
\fl\qquad+(g^2+g+1)(g'(u+\lambda)-2g)(\Psi_+^4+\Psi_-^4)+4(g^2-1)(g'(u+\lambda)-g)(\Psi_+^2+\Psi_-^2)\Bigg],\nn 
\fl \langle F^{ST}_y, F^{ST}_y\rangle= -\frac{(\Psi_+^4-1)}{4g^{\frac32}(u+\lambda)^2}\left((g^2+g+1)(g'(u+\lambda)-2g)(1+\Psi_-^4)+2g'(g^2-1)(u+\lambda)\Psi_-^2\right)u_x\nn
\fl \qquad \frac{1}{16(u+\lambda)^2g^2}\times\Bigg[\left(g^2+g+1)((g'(u+\lambda)-2g)^2+4gf\right)\left(\Psi_+^4+\Psi_-^4\right)\nn
\fl \qquad +4g'(g^2-1)(u+\lambda)(g'(u+\lambda)-2g)(\Psi_+^2+\Psi_-^2)\nn
\fl \qquad +2(u+\lambda)^2(3-g+3g^2)g'^2-8g(g^2-g+1)(u+\lambda)g'\Bigg],\eea\
where for convenience, we denote 
\[ f'=\frac{\partial}{\partial u}{f(u)}, \qquad g'=\frac{\partial}{\partial \lambda }{g(\lambda)}.\]
Similarly, the components of the first fundamental form for the surface induced via generalized symmetries are 
\bea \fl \langle F^{Q}_x, F^{Q}_x\rangle = \frac{Q^2(f''(u+\lambda)^2-2(u+\lambda)f'+2(f-g))^2}{64(u+\lambda)^4g^2}\nn
\fl \qquad \times\left[ (g^2+g+1)(\Psi_+^4+\Psi_-^4)+4f(g^2-1)(\Psi_+^2+\Psi_-^2)+2(g^2-g+3)\right],\nn
\fl \langle F^{Q}_x, F^{Q}_y\rangle  =\frac{Q(f''(u+\lambda)^2-2(u+\lambda)f'+2(f-g))^2)}{32(u+\lambda)^3g^2}\nn
\fl \quad \times\bigg[Qg(g^2+1)+ 2\sqrt{g}(D_xQ-Qu_x)\left((g^2+g+1)(\Psi_+^4-\Psi_-^4)+2(g^2-1)(\Psi_+^2-\Psi_-^2)\right)\nn
\fl \quad +\left(2D_xQu_x-Qf'(u+\lambda)+2gQ\right)\nn
\fl \quad \times \left[(g^2+g+1)(\Psi_+^4+\Psi_-^4)+2(g^2-1)(\Psi_+^2+\Psi_-^2)+2(3g^2-g+3)\right]\Bigg],\nn
\fl \langle F^{Q}_y, F^{Q}_y\rangle  =\frac{(g^2+g+1)}{16g^2(u+\lambda)^2}\nn \fl \quad\times\Bigg[\left[4\sqrt{g}(f'(u+\lambda)+2g)u_xQ^2+2(D_x^2Q)u_x-2(f'(u+\lambda)+2g+2f)Q_xQ\right](\Psi_+^4-\Psi_-^4)\nn
\fl \quad+\left[\left((f'(u+\lambda)+2g)^2+4gf\right)Q^2-4(f'(u+\lambda)+4g)D_xQ Q u_x+4(f+g)(D_xQ)^2\right](\Psi_+^4+\Psi_-^4)\Bigg]\nn
\fl \quad +\frac{(g^2-1)}{4g^{2}(u+\lambda)}\Bigg[2\sqrt{g}\left(f'Q^2u_x+4(u+\lambda)(D_xQ)^2u_x-2(f'(u+\lambda)+2f)D_xQQ\right)(\Psi_+^2-\Psi_-^2)\nn
\fl \quad -\left(4(f'(u+\lambda)+g)D_xQQu_x+(f'(u+\lambda)+2g)f'Q^2+(u+\lambda)f(D_xQ)^2\right)(\Psi_+^2+\Psi_-^2)\Bigg]\nn
\fl \qquad -\frac{(3g^2-g+3)(f'D_xQ Q u_x+f(D_xQ)^2)+g(g^2-g+1)(D_xQ)^2}{g^2}\nn
\fl \qquad +\left(\frac{(3g^2-g+3)f'^2}{8g^2}+\frac{(g^2-g+1)f'}{2g(u+\lambda)}+\frac{(g^2-g+1)(g-f)}{2g(u+\lambda)^2}\right)Q^2.
\eea

For the vector field $\vec{v}_{u_x}$ the components simplify to 
\bea 
\fl \langle F^{Q}_x, F^{Q}_x\rangle = \frac{f(f''(u+\lambda)^2-2(u+\lambda)f'+2(f-g))^2}{64(u+\lambda)^4g^2}\nn
\fl \qquad \times\left[ (g^2+g+1)(\Psi_+^4+\Psi_-^4)+4f(g^2-1)(\Psi_+^2+\Psi_-^2)+2(g^2-g+3)\right],\nn
\fl \langle F^{u_x}_x, F^{u_x}_y\rangle  =\frac{(f''(u+\lambda)^2-2(u+\lambda)f'+2(f-g))^2)}{32 g^{\frac32}(u+\lambda)^3}\nn 
\fl \qquad \times\Bigg[(f'(u+\lambda)-2f)u_x\left((g^2+g+1)(\Psi_+^4-\Psi_-^4)+2(g^2-1)(\Psi_+^2-\Psi_-^2)\right)\nn
\fl \qquad \qquad 2\sqrt{g}f\left((g^2+g+1)(\Psi_+^4+\Psi_-^4)+2(g^2-1)(\Psi_+^2+\Psi_-^2)+2(g^2-g+1)\right)\Bigg],\nn
\fl \langle F^{u_x}_y, F^{u_x}_y\rangle = \frac{(g^2+g+1)(2f-f'(u+\lambda)}{4\sqrt{g}(u+\lambda)^2}(\Psi_+^4-\Psi_-^4)u_x\nn
\fl \qquad+\frac{1}{(u+\lambda)^2g}\Bigg[( f''(u+\lambda)^2-4f'f(u+\lambda)+4f(f+g))(g^2+g+1)(\Psi_+^4+\Psi_-^4)\nn
\fl \qquad \qquad -2( f''(u+\lambda)^2-4f'f(u+\lambda)+4f(f-g))(g^2-g+1)\Bigg].
\eea

\end{document}